\documentclass[12pt,preprint]{aastex}
\usepackage{natbib} 

\def\gtrsim{\lower2pt\hbox{$\buildrel {\scriptstyle >}
   \over {\scriptstyle\sim}$}}
\def\lesssim{\lower2pt\hbox{$\buildrel {\scriptstyle <}
   \over {\scriptstyle\sim}$}}

\slugcomment{Published in ApJ Letters: 679, 37-40 (2008)}

\shorttitle{Spin of Black Hole M33 X-7}
\shortauthors{Liu et al.}

\begin{document}

\title{Precise Measurement of the Spin Parameter of the Stellar-Mass
Black Hole M33 X-7}

\author{Jifeng Liu\altaffilmark{1}, Jeffrey E.\
        McClintock\altaffilmark{1}, Ramesh Narayan\altaffilmark{1},
        Shane W.\ Davis\altaffilmark{2}, and Jerome A.\
        Orosz\altaffilmark{3}}

\altaffiltext{1}{Harvard-Smithsonian Center for Astrophysics,60 Garden Street,
Cambridge, MA 02138} \altaffiltext{2}{Institute for Advanced Study, Einstein
Drive, Princeton, NJ 08540} \altaffiltext{3}{Department of Astronomy, San Diego
State University, 5500 Campanile Drive, San Diego, CA 92182}

\begin{abstract}

In prior work, {\it Chandra} and Gemini-North observations of the eclipsing
X-ray binary M33 X-7 have yielded measurements of the mass of its black hole
primary and the system's orbital inclination angle of unprecedented accuracy.
Likewise, the distance to the binary is known to a few percent.  In an analysis
based on these precise results, fifteen {\it Chandra} and {\it XMM-Newton}
X-ray spectra, and our fully relativistic accretion disk model, we find that
the dimensionless spin parameter of the black hole primary is $a_* = 0.77 \pm
0.05$.  The quoted 1-$\sigma$ error includes all sources of observational uncertainty.  Four {\it
Chandra} spectra of the highest quality, which were obtained over a span of
several years, all lead to the same estimate of spin to within statistical
errors (2\%), and this estimate is confirmed by 11 spectra of lower quality.
There are two remaining uncertainties: (1) the validity of the relativistic
model used to analyze the observations, which is being addressed in ongoing
theoretical work; and (2) our assumption that the black hole spin is
approximately aligned with the angular momentum vector of the binary, which can
be addressed by a future X-ray polarimetry mission.

\end{abstract}

\keywords{Galaxies: individual(M33) --- X-rays: binaries --- black hole
physics --- binaries: individual (M33 X-7)}

\section{INTRODUCTION}

M33 X-7 is the first stellar-mass black hole to be discovered that is
eclipsed by its companion \citep{pie06}.
The X-ray eclipse and
the precisely known distance of this system, $D=840\pm20$ kpc, underpin
the most accurate dynamical model that has been achieved for any of the
21 known black hole binaries (\citealt{oro07}, hereafter O07).  The two dynamical
parameters of interest in this Letter are the black hole mass $M =
15.65\pm1.45 M_\odot$ and the orbital inclination angle $i = 74.6^\circ
\pm 1.0^\circ$ (O07).

Our group has published spin estimates for three stellar-mass black
holes using the X-ray continuum fitting method: GRO J1655-40, $a_* =
0.65-0.75$; 4U 1543-47, $a_* = 0.75-0.85$; and GRS 1915+105, $a_* =
0.98-1.0$ (\citealt{sha06}, hereafter S06; \citealt{mcc06}, hereafter M06).  For LMC X-3, \citet{dav06} find $a_* < 0.26$.
Meanwhile, the Fe line method has been used to obtain two additional
estimates of black hole spin (\citealt{bre06}; \citealt{mil08}).
The dimensionless spin parameter $a_* \equiv a/M = cJ/GM^2$,
where $M$ and $J$ are the mass and angular momentum of the black hole;
$-1 \leq a_* \leq 1$ \citep{sha86}.

The continuum-fitting method, which was pioneered by \citep{zha97} (also see \citealt{gie01}),
is based on the existence of
an innermost stable circular orbit (ISCO) for a particle orbiting a
black hole, inside which the particle suddenly plunges into the hole.
In the continuum-fitting method, one identifies the inner edge of the
black hole's accretion disk with the ISCO and estimates the radius
$R_{\rm ISCO}$ of this orbit by fitting the X-ray continuum spectrum.
Since the dimensionless radius $r_{\rm isco} \equiv R_{\rm
ISCO}/(GM/c^2)$ is solely a monotonic function of the black hole spin
parameter \citep{sha86}, knowing its value allows one to
immediately infer the black hole spin parameter $a_*$.

Our estimates of spin are based on our fully relativistic accretion disk
model \citep{lli05} and an advanced treatment of spectral hardening
\citep{dav05}.  We consider only rigorously-selected
thermal-state data \citep{rem06}, which are largely
free of the effects of Comptonization.  Furthermore, we only accept data
for which the bolometric disk luminosity is moderate, $L/L_{\rm Edd} <
0.3$, in order to ensure 
that the standard geometrically-thin thermal disk
model is applicable (\citealt{sha08}; M06).

For the continuum-fitting method to succeed, it is essential to have
accurate values of the black hole mass, orbital inclination and distance
(M06), quantities that are known precisely in the case of M33 X-7.
Other virtues of M33 X-7 for the determination of spin, are the
abundance of {\it Chandra} and {\it XMM} data, the remarkably thermal
and featureless spectrum of the X-ray source, and its moderate
luminosity (\S3).

\section{DATA SELECTION AND REDUCTION}

There have been 17 {\it Chandra} ACIS observations and 12 {\it
XMM-Newton} EPIC observations of M33 X-7.  We analyzed {\it Chandra}
observations (downloaded from the {\it Chandra} Data Archive) with CIAO
3.4 and extracted the spectra from source
ellipses enclosing 95\% of the source photons as reported by {\sc
wavdetect}.  The {\it XMM} observations were downloaded from the HEASARC
archive and analyzed with SAS 7.0.0 \citep{gab04}.  Using the
standard procedures and excluding intervals of high background, we
extracted separate spectra from the PN and two MOS chips using radii of
400 pixels (i.e., $20^{\prime\prime}$) and fitted them independently.


The resultant count rate data were folded on the M33 X-7 X-ray eclipse
ephemeris \citep{pie06}: HJD $(2453639.119 \pm 0.005) \pm
N(3.453014 \pm 0.000020)$.  The folded light curve for all the {\it
Chandra} data are shown in Figure $2a$ in O07.  For the purpose of
measuring the spin of M33 X-7, we excluded (1) spectra obtained in the
phase range --0.3 to 0.2, i.e., during the eclipse or the pre-eclipse
period of erratic X-ray variability (which is presumably caused by the
accretion stream; O07); (2) four ACIS spectra (ObsIDs 6384, 6385, 7170,
7171) that are severely affected by pile-up; and (3) all spectra that
contain less than 1000 counts.  The 15 spectra so selected are listed in
Table 1 and comprise eight {\it Chandra} ACIS spectra and seven EPIC
PN/MOS spectra from five {\it XMM} observations.  We refer throughout to
the four ACIS spectra with $\gtrsim$ 5,000 counts as the ``gold''
spectra and to the rest as the ``silver'' spectra.

\section{ANALYSIS AND RESULTS}

The procedures used here are precisely the same as those that are described
fully in M06.  Briefly, the relativistic accretion disk model {\sc kerrbb2} has
just two fit parameters, namely the black hole spin $a_*$ and the mass
accretion rate $\dot M$ (or equivalently, $a_*$ and the Eddington-scaled
bolometric luminosity, $l \equiv L_{\rm bol}(a_*,\dot M)/L_{\rm Edd}$; M06).
In the case of M33 X-7, we also fit for a third parameter, $N_{\rm H}$, the
hydrogen column density ({\sc phabs} in XSPEC).

The spectral hardening factor $f \equiv T_{\rm col}/T_{\rm eff}$ was
computed as a function of $l$ for the appropriate metallicity of M33 X-7
($Z = 0.1 Z_\odot$; O07) using the model of \citealt{dah06} ({\sc
bhspec} in XSPEC).  These values of $f$ are contained in a pair of
lookup tables, which correspond to two representative values of the
viscosity parameter ($\alpha = 0.01,0.1$; M06) for a wide range of the
spin parameter (e.g., $0 < a_* < 0.99$).  We find
that our results are quite insensitive to the choice of $\alpha$ or an
increase in metallicity.  We have also experimented with varying the
input parameters $M$, $D$, $i$ and $N_{\rm H}$, and we find that the
values of $f$ are scarcely affected.

All of the spectra were well-fitted using a simple absorbed {\sc
kerrbb2} model [i.e., {\sc phabs(kerrbb2)} in XSPEC].  Notably, neither
Fe line/edge components nor an additional nonthermal component was
required, as they were in our earlier work (R06, M06).  We fitted each
spectrum for $a_*$, the mass accretion rate $\dot M$, and the neutral
hydrogen column density $N_{\rm H}$ with the input parameters fixed at
their baseline values (see \S1; O07).  The normalization was fixed at
unity (as appropriate when $M$, $i$ and $D$ are held fixed).  We
included the effects of limb darkening (lflag = 1) and returning
radiation effects (rflag = 1), and we set the torque at the inner
boundary of the accretion disk to zero ($\eta=0$).

The fits obtained for all 15 spectra are quite acceptable with
$\chi^2_\nu < 1.2$; results for fits over the energy range 0.3--8 keV
are summarized in Table 1.  An inspection of the fitting residuals for
all the spectra show them to be free of any systematic effects, as
illustrated in Figure~1.  Given the modest luminosities, $0.07 < l <
0.11$ (Table 1), which are well below our selection limit of $l=0.3$,
the accretion disk in M33 X-7 is quite thin, $H/R \le 0.04$ (see Fig.\
17 in \citealt{sha08}), and our assumption of zero torque at the
inner boundary is likely to be valid.

Figure 2 shows plots of $a_*$ for all 15 observations, which are ordered
by the number of counts detected.  Each of the four panels corresponds
to a different choice for the energy interval used in fitting the data
(e.g., 0.3--8 keV, 0.5--8 keV, etc.); a comparison of the results in the
four panels shows that this choice is quite unimportant.  The four gold
spectra with $\gtrsim\ 5,000$ counts each (solid symbols) yield spin
estimates that agree with their mean value (indicated by the dotted
lines) typically to within their $\approx 2$\% statistical
uncertainties.  The stability of these four gold spectra is especially
remarkable given that three of the observations were separated by
3-month intervals in 2005--2006, and one of them was obtained five years
earlier in 2000 (Table~1).  The dispersion for the 11 silver spectra
that have $\lesssim\ 3,000$ counts (open symbols) is much larger.
However, in each panel, the mean of these 11 spin values agrees with the
mean determined using the gold spectra to within $\approx 1$\%.  As
concluded in the caption of Figure 2, our adopted average spin for the
four gold spectra is $\bar{a}_* = 0.77$ with a standard deviation of
$\Delta a_{*} = 0.02$.

In order to determine the error in $a_*$ due to the combined
uncertainties in $M$, $i$ and $D$ (\S1), we performed Monte Carlo
simulations assuming that the uncertainties in these parameters are
normally and independently distributed.  The results for 3,000
simulation runs are plotted in Figure 3.  The histogram of
$\bar{a}_*-\bar{a}_{*0}$ shows that the $1\sigma$ error in the spin due
to the combined uncertainties of the three input parameters is about
$\Delta a_* = 0.05$.  The error is dominated by the uncertainty in $M$;
the uncertainties in $i$ and $D$ are relatively unimportant.  This error
is based on a readily available table that was computed for solar
metallicity.  Despite this limitation, we believe that our error
estimate is accurate because the effects of going from $Z=0.1Z_{\odot}$
to $Z=Z_{\odot}$ are very small at the luminosities in question, $l
\approx 0.1$. 

\section{DISCUSSION}

The largest error in our spin estimate arises from the uncertainties in
the validity of the disk model we employ.  For example, the spin depends
on accurate model determinations of the hardening factor $f$; this
problem is quite tractable and vigorous theoretical efforts are underway
(\citealt{dav05, dav06, bla06}).  Possibly
more problematic is our assumption that the viscous torque vanishes at
the ISCO and that there is no significant emission from the gas inside
the ISCO.

Hydrodynamic models of the accretion disk indicate that the viscous torque at
the ISCO as well as emission from inside the ISCO should both be negligible for
the geometrically thin disks and low luminosities ($l\le0.3$) that we restrict
ourselves to (\citealt{afs03}; S08).  The emission from inside the
ISCO causes rather modest errors in spin estimates; in the case of M33 X-7, the
estimated error is $\Delta a_* \le 0.01$ since $l\le0.1$ and hence $R/H \le
0.04$ (M06; S08).  On the other hand, MHD simulations of accretion flows around
black holes \citep{haw02, bec08} find a large torque
at the ISCO and substantial dissipation inside the ISCO.  We note, however,
that these simulations carried out so far are for geometrically thick systems,
with $H/R\sim 0.2$; these flows are nearly an order of magnitude thicker than
the disk in M33 X-7.  In the hydrodynamic models of \citet{sha08} the
stress at the ISCO increases rapidly with increasing disk thickness, so it is
conceivable that there is no serious disagreement between the hydrodynamic and
MHD results.  Numerical MHD simulations of truly thin disks are necessary to
resolve this issue.  We note that a recent MHD simulation of a geometrically
thin accretion disk for a pseudo-Newtonian potential does show a dramatic drop
in the mid-plane density and vertical column density over a narrow range of
radii close to the ISCO \citep{rey08}.

Although there is theoretical uncertainty about conditions near the ISCO, there
is a long history of evidence suggesting that fitting the X-ray continuum is a
promising approach to measuring black hole spin.  This history begins in the
mid-1980s with the simple non-relativistic multicolor disk model \citep{mit84}, which returns the color temperature $T_{\rm in}$ at the inner-disk
radius $R_{\rm in}$. \citet{tan95} summarize examples of the steady
decay (by factors of 10--100) of the thermal flux of transient sources during
which $R_{\rm in}$ remains quite constant (see their Fig.\ 3.14).  More
recently, this evidence for a constant inner radius in the thermal state has
been presented for a number of sources in several papers via plots showing that
the bolometric luminosity of the thermal component is approximately
proportional to $T^4$ (\citealt{mcc07}, and references therein).
Obviously, this non-relativistic analysis cannot provide a secure value for the
radius of the ISCO nor even establish that this stable radius is the ISCO.
Nevertheless, the presence of a fixed radius indicates that the
continuum-fitting method is a well-founded approach to measuring black hole
spin.

It is reasonable to assume that the inner X-ray-emitting portion of the disk is
aligned with the spin axis of the black hole by the Bardeen-Petterson effect
\citep{lod05}.  Throughout, in making use of the orbital inclination angle, we
have assumed that the black hole spin is aligned with the angular momentum
vector of the binary system.  As Figure~3 indicates, if any misalignment is
$\lesssim\ 3^\circ$, then it will contribute an error in $a_*$ that is no
larger than our total observational error of $\Delta a_* = 0.05$.  There is no
evidence for significant misalignments despite the often-cited examples of GRO
J1655-40 and SAX J1819.3-2525 (see \S2.2 in \citealt{nar05}; but see
\citealt{mac02}).  The clear-cut way to assess the degree of alignment is via
X-ray polarimetric observations of black hole systems in the thermal state (Li
et al.\ 2008, in preparation).

What is the origin of the spin of M33 X-7?  Was the black hole born with its
present spin, or was it torqued up gradually via the accretion flow supplied by
its companion?  In order to achieve a spin of $a_* = 0.77$ via disk accretion,
an initially non-spinning black hole must accrete $4.9 M_\odot$ from its donor
\citep{kin99} in becoming the $M = 15.65 M_\odot$ that we observe today
(O07).  However, to transfer this much mass even in the case of
Eddington-limited accretion ($\dot{M}_{\rm Edd} \equiv L_{\rm Edd}/c^2 \approx
4\times10^{-8} M_\odot/ {\rm yr}$) requires $\sim 120$ million years, whereas
the age of the system is only 2--3 million years (O07).  Thus, it appears that
the spin of M33 X-7 must be natal, which is the same conclusion that has been
reached for two other stellar black holes (S06, M06; but see \citealt{beth03}
on the possibility of hypercritical accretion)

M33 X-7's secure dynamical data and distance, the X-ray source's clean
thermal-state spectrum and moderate luminosity, and an abundance of {\it
Chandra} and {\it XMM} data have provided arguably the most secure
estimate of black hole spin that has been achieved to date: $a_* = 0.77
\pm 0.05$, where the error estimate includes all sources of
observational error.  Since an astrophysical black hole can be described by
just the two parameters that specify its mass and spin \citep{sha86}, we now
have a {\it complete} description of an asteroid-size object that is situated
at a distance of about one Mpc.

\acknowledgements

JFL and SWD acknowledge support from NASA through the {\it Chandra}
Fellowship Program, grants PF6-70043 and PF6-70045.  JEM acknowledges
support from NASA grant AR8-9006X.  We thank Rebecca Shafee for
technical advice and Jack Steiner for critical comments on the
manuscript.

\title{\sc Erratum: ``Precise Measurement of the Spin Parameter of the Stellar-mass Black Hole M33 X-7'' (ApJL, 679, 37L [2008])}

In the paper ``Precise Measurement of the Spin Parameter of the
Stellar-Mass Black Hole M33 X-7'' by Jifeng Liu, Jeffrey E.\
McClintock, Ramesh Narayan, Shane W.\ Davis, and Jerome A.\ Orosz
(ApJ, 679, L37 [2008]), the reported value of the black-hole spin
parameter $a_*=0.77\pm0.05$ is in error.  The correct value is larger
by 0.068 and is $a_*=0.84\pm0.05$.  The error is the result of a bug
in the XSPEC accretion-disk model {\sc
kerrbb}.\footnote{http://heasarc.gsfc.nasa.gov/docs/xanadu/xspec/issues/archive/issues.12.5.0an.html
(patch 12.5.0a).}  Prior to 1 December 2008, the model's two parameter
flags that switch limb darkening and self-irradiation of the disk
on/off were reversed (e.g., ``par8'' incorrectly controlled limb
darkening rather than self-irradiation).  In computing tables of the
spectral hardening factor $f$, we use both {\sc kerrbb} and the disk
atmosphere model {\sc bhspec} \citep{mcc06}.  Because the
latter model does not include the effect of self-irradiation, we
switch this feature off in {\sc kerrbb} when computing the $f$-tables.
In this instance, because of the bug we switched off limb darkening
instead of self-irradiation, which corrupted our results.  Meanwhile,
our earlier spin results for GRS 1915+105 (McClintock et al.\ 2006)
and for 4U 1543--47 and GRO J1655--40 \citep{sha06} are
unaffected by the bug.

The figures and tabular data in the paper are essentially unaffected,
apart from the increase in $a_*$ and corresponding decreases in $f$ and
the Eddington-scaled luminosity $l$ (8.7\% and 4.5\%, respectively, for
the four gold spectra).  The higher spin increases somewhat our estimate
of how much mass ($4.9~M_{\odot}$) and time ($\sim120$ million years)
would be required to spin up an initially nonspinning black hole to the
present spin of M33 X-7. In order to achieve $a_*=0.84$, the black hole
must accrete $5.7~M_{\odot}$, which would require $\sim140$ million
years.  Because the age of the binary system is only 2--3 million years
this change does not at all affect our conclusion that the spin of M33
X-7 is natal.

\bibliography{natms}

\begin{deluxetable}{rrrrcccrrr}

\tabletypesize{\scriptsize}

\tablecaption{{\sc Kerrbb2} fit results for M33 X-7 in 0.3-8
keV\tablenotemark{a}}

\tablehead{\colhead{spectrum} & \colhead{obs-date} & \colhead{Texp} & \colhead{counts} & \colhead{$a_*$} &  
\colhead{$\dot{M}$} & \colhead{$n_{\rm H}$} & \colhead{$f_{\rm col}$} & \colhead{$\lg l$} & \colhead{$\chi^2_\nu$/dof} 
}

\startdata

acis6376     & 2006-03-03 & 93.1 & 9748 & 0.751 $\pm$ 0.026 & 1.88 $\pm$ 0.12 & 11.1 $\pm$ 1.1 & 1.78 & -1.01 & 1.07/180 \\
acis6387     & 2006-06-26 & 77.3 & 7271 & 0.782 $\pm$ 0.019 & 1.64 $\pm$ 0.10 & 11.4 $\pm$ 1.2 & 1.76 & -1.05 & 0.93/157 \\
acis6382     & 2005-11-23 & 72.3 & 6515 & 0.772 $\pm$ 0.030 & 1.72 $\pm$ 0.14 &  9.9 $\pm$ 1.4 & 1.78 & -1.04 & 1.17/152 \\
acis1730     & 2000-07-12 & 49.5 & 4855 & 0.800 $\pm$ 0.026 & 1.37 $\pm$ 0.11 &  6.1 $\pm$ 1.3 & 1.77 & -1.12 & 1.15/126 \\
\hline                    
acis7344     & 2006-07-01 & 21.5 & 1711 & 0.873 $\pm$ 0.031 & 1.05 $\pm$ 0.14 &  8.6 $\pm$ 2.8 & 1.75 & -1.16 & 0.69/55  \\
acis6386     & 2005-10-31 & 14.9 & 1491 & 0.786 $\pm$ 0.041 & 1.55 $\pm$ 0.21 & 12.8 $\pm$ 3.2 & 1.77 & -1.07 & 0.95/49  \\
acis7197     & 2005-11-03 & 12.7 & 1117 & 0.892 $\pm$ 0.043 & 0.97 $\pm$ 0.20 &  9.1 $\pm$ 4.1 & 1.75 & -1.17 & 0.99/37  \\
acis7208     & 2005-11-21 & 11.5 & 1014 & 0.678 $\pm$ 0.110 & 1.73 $\pm$ 0.39 & 13.9 $\pm$ 4.6 & 1.78 & -1.10 & 0.81/33  \\
\hline                    
PN0102642301 & 2002-01-27 & 10.0 & 2724 & 0.832 $\pm$ 0.031 & 1.34 $\pm$ 0.14 &  7.3 $\pm$ 1.0 & 1.75 & -1.09 & 0.84/103 \\
PN0102641201 & 2000-08-02 & 10.3 & 1836 & 0.618 $\pm$ 0.056 & 2.51 $\pm$ 0.29 & 12.0 $\pm$ 1.5 & 1.78 & -0.97 & 0.87/69  \\
PN0141980801 & 2003-02-12 & 8.4  & 1596 & 0.636 $\pm$ 0.074 & 2.50 $\pm$ 0.36 & 10.7 $\pm$ 1.5 & 1.78 & -0.96 & 0.90/60  \\
PN0141980601 & 2003-01-23 & 11.6 & 1545 & 0.656 $\pm$ 0.077 & 2.21 $\pm$ 0.32 & 12.5 $\pm$ 1.6 & 1.77 & -1.00 & 1.00/59  \\
M10102642301 & 2002-01-27 & 12.3 & 1199 & 0.841 $\pm$ 0.042 & 1.30 $\pm$ 0.20 &  6.7 $\pm$ 2.1 & 1.75 & -1.10 & 0.89/40  \\
PN0102640401 & 2000-08-02 & 9.2  & 1136 & 0.838 $\pm$ 0.039 & 1.42 $\pm$ 0.20 &  6.8 $\pm$ 1.8 & 1.75 & -1.06 & 1.15/44  \\
M20102642301 & 2002-01-27 & 12.3 & 1135 & 0.839 $\pm$ 0.043 & 1.25 $\pm$ 0.20 &  7.2 $\pm$ 2.2 & 1.75 & -1.12 & 0.83/39  \\

\enddata

\tablenotetext{a}{The columns are (1) ID number; (2) date of observation; (3)
exposure time in ksec; (4) no. of counts; (5) spin parameter; (6) mass
accretion rate in $10^{18}$ g s$^{-1}$; (7) hydrogen column density in
$10^{20}$ cm$^{-2}$; (8) spectral hardening factor; (9) Eddington-scaled
bolometric luminosity; and (10) reduced $\chi^2$ per dof. }

\end{deluxetable}

\clearpage


\begin{figure}
\includegraphics[width=7.0in, height=7.0in, angle=270]{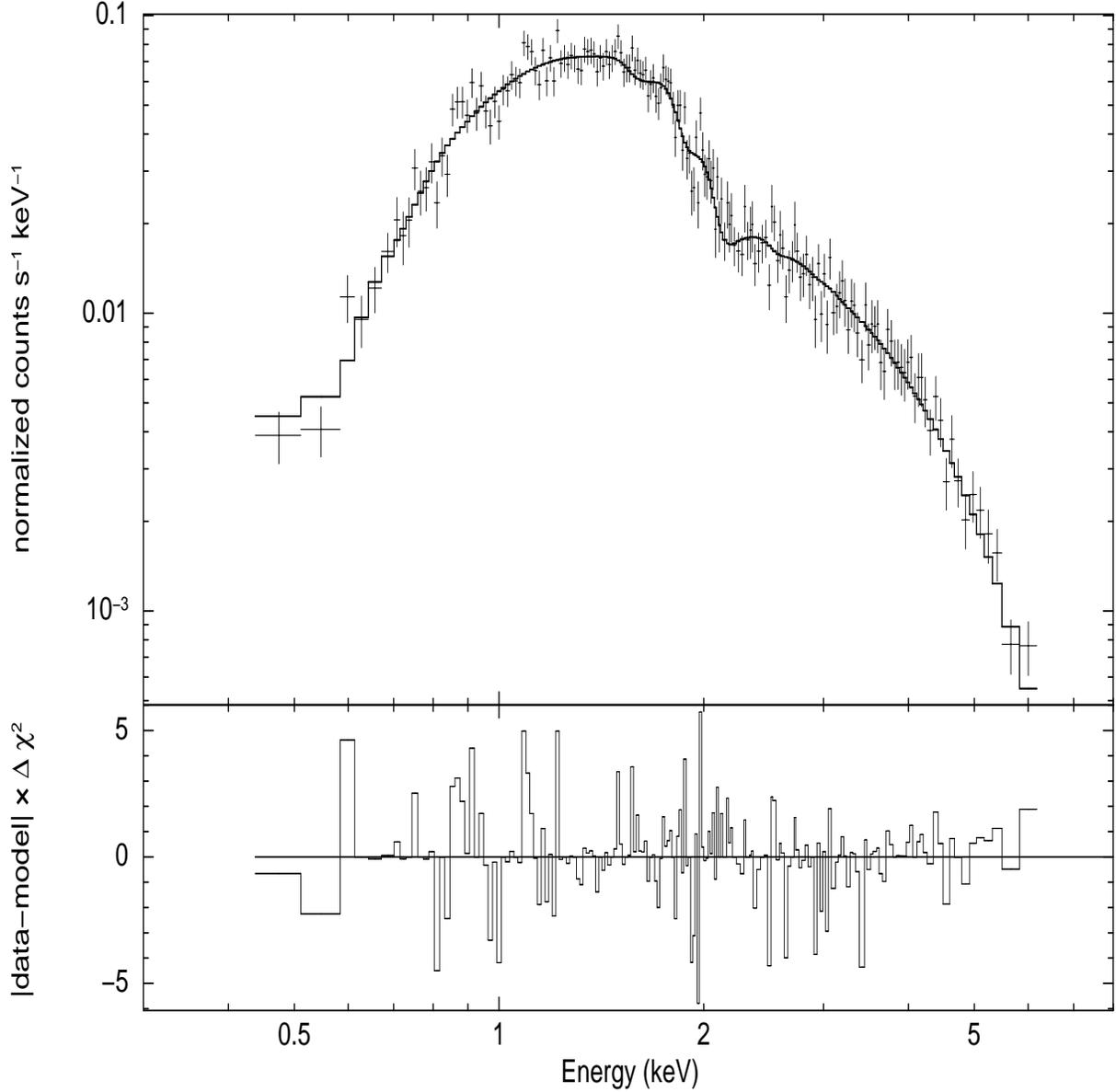}
\caption{X-ray spectrum of M33 X-7.  ({\it upper panel}) This spectrum
(ObsID 6376) is representative of the four gold spectra (see text).  The
histogram shows a model that has been fitted to the spectrum (0.3--8.0
keV), which is comprised of only the thermal disk component ({\sc
kerrbb2}) and a low-energy absorption model ({\sc phabs}).  The fit
parameters are summarized in Table~1.  ({\it lower panel}) The fit is
good ($\chi_\nu^2/dof = 1.074/180$) and the fit residuals show no
systematic structure; in particular, there is no evidence for an
Fe-line, absorption edges, or a nonthermal power-law/Comptonization
component of emission at higher energies.}
\end{figure}


\begin{figure}
\plotone{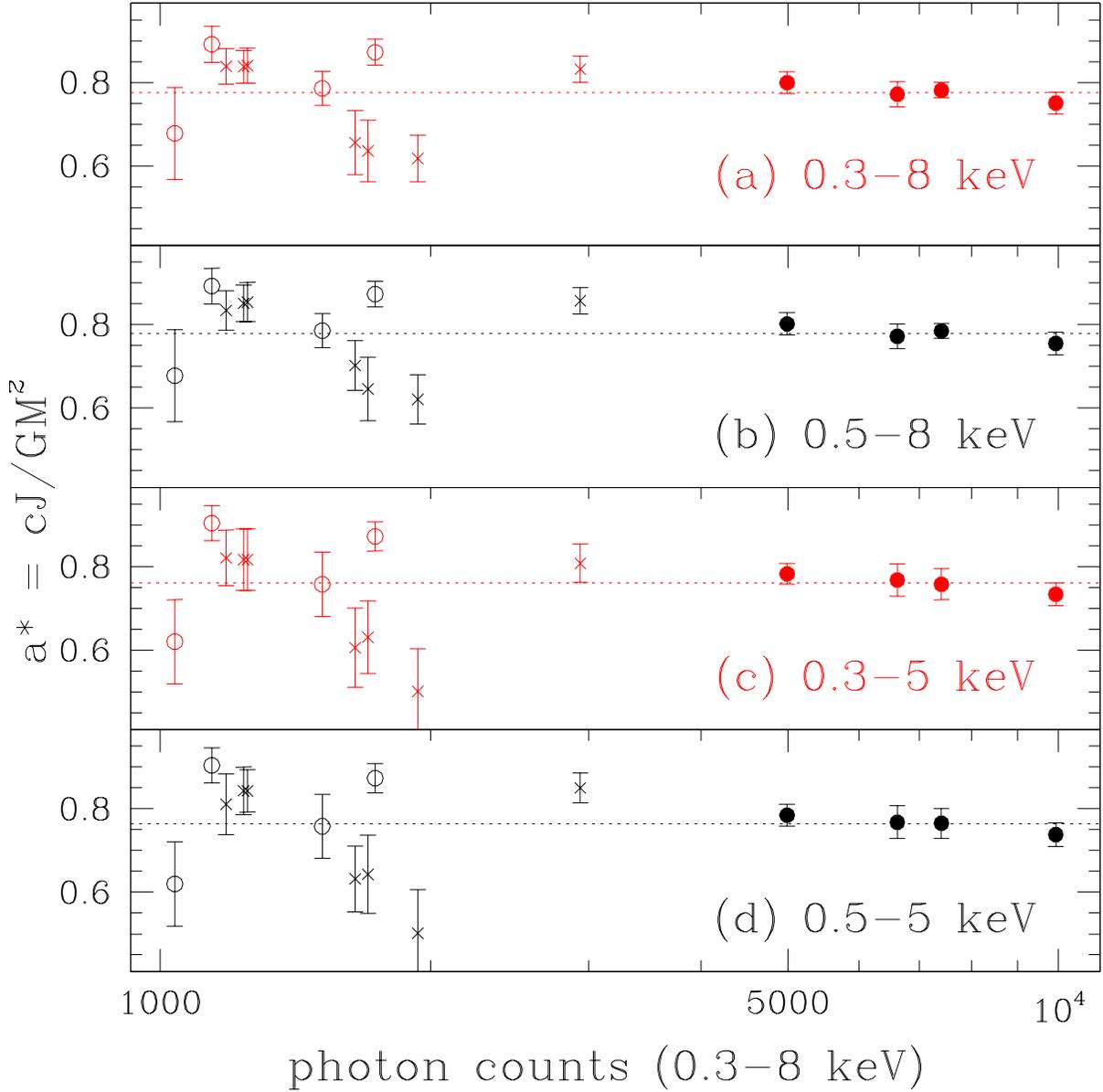}
\caption{Spin results for all 15 spectra ordered by total 0.3--8 keV
counts.  ({\it a}) Results based on spectral fits over the energy
interval 0.3--8 keV.  Filled circles are for the gold {\it Chandra}
spectra, and the other plotting symbols are for the 11 silver spectra,
which include both {\it Chandra} ACIS spectra (open circles) and {\it
XMM-Newton} EPIC spectra (crosses).  The indicated uncertainties are at
the 90\% level of confidence.  The dotted line indicates the average
spin for the 4 gold spectra, $\bar{a}_* = 0.776 \pm 0.018$.  The average
for the 11 silver spectra is almost identical, although the dispersion
is much greater, $\bar{a}_* = 0.772 \pm 0.098$. ({\it b}--{\it d}) Same
as panel {\it a} except the fit interval is as indicated rather than
0.3--8 keV.}
\end{figure}

\begin{figure}
\plotone{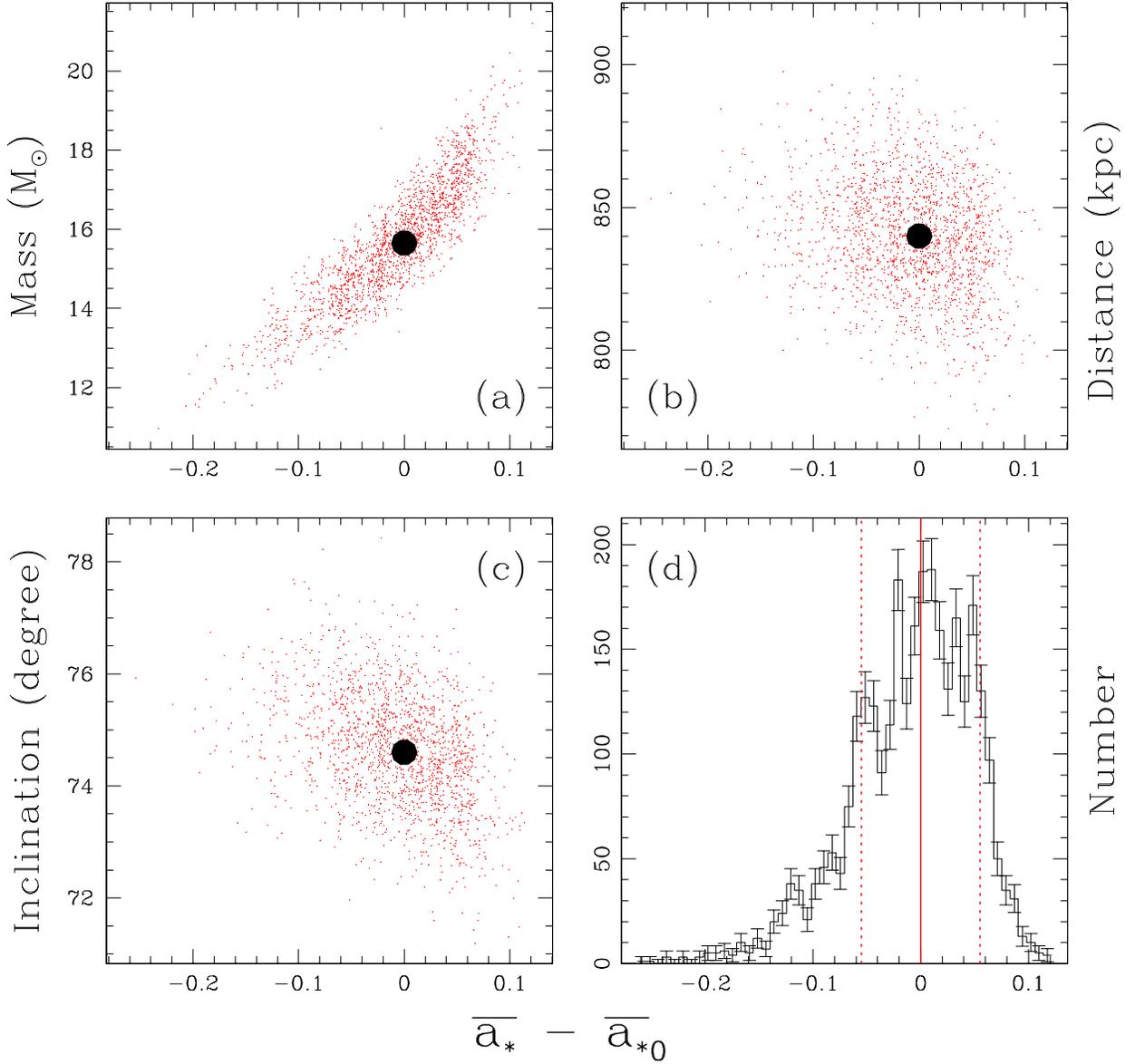}
\caption{Effect on the spin parameter $a_*$ of varying the input
parameters $M$, $i$ and $D$.  (a) Spin versus mass $M$ for 3,000 sets of
parameters drawn at random. The black filled circle 
indicate our final adopted estimate of the spin ($\bar{a}_{*0} = 0.77
\pm 0.02$) for $M$, $i$ and $D$ at their baseline values.  (b) Spin
versus distance $D$.  (c) Spin versus inclination angle $i$.  (d)
Histogram of spin displacements for 3,000 parameters sets.  The vertical
solid line indicates the average spin ($\bar{a}_* = 0.77$). The two
dotted lines enclose 68.3\% of the spin values centered on the solid
line; the half-separation, $\Delta(\bar{a}_*-\bar{a}_{*0}) = 0.053$,
represents the $1\sigma$ error in the average spin of $\bar{a}_{*0} =
0.77$.}
\end{figure}

\end{document}